\documentclass[final,5p,times,twocolumn]{elsarticle}

\usepackage{lineno,hyperref}

\journal{Nucl. Instrum. Methods Phys. Res. A (NIMA Proc. SI: RICH 2018)}

\begin{document}

\begin{frontmatter}

\title{Developing a silica aerogel radiator for the HELIX ring-imaging Cherenkov system}


\author[address-a]{Makoto Tabata\corref{mycorrespondingauthor}}
\cortext[mycorrespondingauthor]{Corresponding author}
\ead{makoto@hepburn.s.chiba-u.ac.jp }

\author[address-b]{Patrick Allison}
\author[address-b]{James J. Beatty}
\author[address-c]{Stephane Coutu}
\author[address-d]{Mark Gebhard}
\author[address-e]{Noah Green}
\author[address-f]{David Hanna}
\author[address-g]{Brandon~Kunkler}
\author[address-d]{Mike Lang}
\author[address-b]{Keith McBride}
\author[address-c]{Isaac Mognet}
\author[address-h]{Dietrich M\"{u}ller}
\author[address-d]{James Musser}
\author[address-i]{Scott Nutter}
\author[address-j]{Nahee Park}
\author[address-e]{Michael Schubnell}
\author[address-e]{Gregory Tarl\'{e}}
\author[address-e]{Andrew Tomasch}
\author[address-g]{Gerard Visser}
\author[address-h]{Scott P. Wakely}
\author[address-h]{Ian Wisher}

\address[address-a]{Department of Physics, Chiba University, Chiba, Japan}
\address[address-b]{Department of Physics, Ohio State University, Columbus, OH, USA}
\address[address-c]{Department of Physics, Pennsylvania State University, University Park, PA, USA}
\address[address-d]{Department of Physics, Indiana University Bloomington, Bloomington, IN, USA}
\address[address-e]{Department of Physics, University of Michigan, Ann Arbor, MI, USA}
\address[address-f]{Department of Physics, McGill University, Montreal, Canada}
\address[address-g]{Center for Exploration of Energy and Matter (CEEM), Indiana University Bloomington, Bloomington, IN, USA}
\address[address-h]{Department of Physics, University of Chicago, Chicago, IL, USA}
\address[address-i]{Department of Physics and Geology, Northern Kentucky University, Highland Heights, KY, USA}
\address[address-j]{Wisconsin IceCube Particle Astrophysics Center (WIPAC), University of Wisconsin--Madison, Madison, WI, USA}

\begin{abstract}
This paper reports the successful fabrication of silica aerogel Cherenkov radiators produced in the first batches from a 96-tile mass production performed using pin-drying technique in our laboratory. The aerogels are to be used in a ring-imaging Cherenkov detector in the spectrometer of a planned balloon-borne cosmic-ray observation program, HELIX (High Energy Light Isotope eXperiment). A total of 36 transparent, hydrophobic aerogel tiles with a high refractive index of 1.16 and dimensions of 10 cm $\times $ 10 cm $\times $ 1 cm will be chosen as the flight radiators. Thus far, 40 out of the 48 tiles fabricated were confirmed as having no tile cracking. In the first screening, 8 out of the first 16 tiles were accepted as flight-qualified candidates, based on basic optical measurement results. To fit the aerogel tiles into a radiator support structure, the trimming of previously manufactured prototype tiles using a water-jet cutting device was successful.
\end{abstract}


\end{frontmatter}


\section{Introduction}

We have been developing silica aerogels for use as radiators in a ring-imaging Cherenkov (RICH) detector to be used in the spectrometer of the High Energy Light Isotope eXperiment (HELIX) \cite{cite1,cite2}. The HELIX program is a balloon-borne cosmic-ray experiment designed to measure the mass of light cosmic-ray isotopes, in particular those of beryllium. The main objective is to explore the propagation mechanism of cosmic rays by measuring the relative abundances of key light cosmic-ray isotopes based on precise particle identification. The first flight is scheduled during the National Aeronautics and Space Administration's 2020/21 Antarctic long-duration balloon campaign.

The HELIX instrument consists of a 1 T superconducting magnet, a gas drift chamber, time-of-flight (ToF) counters, and an aerogel RICH system. The HELIX RICH is used for measuring the velocity of cosmic-ray isotopes over an energy range from 1 to 10 GeV/n. The system is a proximity-focusing RICH with an expansion length of 50 cm, and a focal-plane composed of a silicon photomultiplier module array with a pixel size of 6 mm $\times $ 6 mm. Hydrophobic aerogels with a nominal refractive index, $n$, of 1.15 were chosen as the radiators based on results obtained during their engineering production.

Developing highly transparent aerogels with an ultrahigh index exceeding 1.10 without cracking is a difficult challenge. In late 2016, we began the experimental fabrication of aerogels specifically for use in HELIX. We determined that the production of $n$ = 1.15 tiles is feasible using pin-drying technology \cite{cite3}. It is worth noting that others have previously developed hydrophilic aerogels where $n$ = 1.13 by a sintering technique \cite{cite4}. In early 2018, we succeeded in prototyping highly transparent (transmittance of approximately 73\% at 400 nm wavelength), 10 mm thick aerogels, which were close to being flight-qualified radiators. By the end of March 2018, we had begun mass production of 96 tiles as candidates for flight aerogels. Here we report the progress in mass production and the basic optical characterization results (i.e., the refractive index measured at the corners of tiles and transmission length) from the early batches of the ultrahigh-index hydrophobic aerogels.

\begin{figure}[t]
\centering 
\includegraphics[width=0.38\textwidth,keepaspectratio]{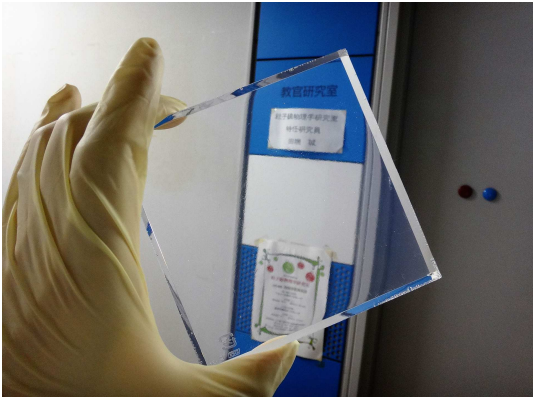}
\caption{The first highly transparent aerogel tile obtained from mass production.}
\label{fig:fig1}
\end{figure}

\section{Requirements for aerogel Cherenkov radiators}

The staging approach of the HELIX program requires two different refractive indices from the aerogel radiators. The first stage/flight demands $n$ = 1.12--1.18 in order to cover the energy range from 1 to 3 GeV/n, which is within the scope of the present study. In the future, second stage aerogels with $n \sim $ 1.03 will be stacked on the radiator module to cover the energy range up to 10 GeV/n. We rejected aerogels with $n$ = 1.18 because the effective mass production of higher index tiles using pin-drying technology is challenging. Feasibility studies for systematically fabricating aerogels with $n$ = 1.12--1.15 were successful; thus, we chose $n$ = 1.15 as the index for the flight aerogels. This index will provide better velocity resolution and an opportunity for cross-calibrating the velocity measurement systems (i.e., RICH and ToF) around 1 GeV/n.

The aerogel radiator module was designed to have dimensions of 60 cm $\times $ 60 cm, which is segmented into 36 blocks, each with dimensions 10 cm $\times $ 10 cm. The production of aerogel tiles with a lateral dimension of approximately 12 cm and $n \sim $ 1.15 is feasible. Tiles with dimensions of 10 cm $\times $ 10 cm can be prepared by trimming the tile edges of the larger raw tiles using a water-jet cutting machine. At least 36 tiles with no cracking are required to fill the radiator module, which is within the scope of the laboratory-scale mass production at Chiba University. The tile thickness was determined to be 1 cm to minimize Cherenkov photon's emission point uncertainties along the cosmic-ray particle trajectories in the tile thickness direction. Despite being rather thin radiators, sufficient photon yield (greater than 140 photoelectrons for the particles with a charge of 4 and velocity, $\beta $, of 1 \cite{cite5}) can be expected from the tiles due to the high index of the radiators and the fact that the measured particles have charges greater than 1.

\begin{figure}[t]
\centering 
\includegraphics[width=0.43\textwidth,keepaspectratio]{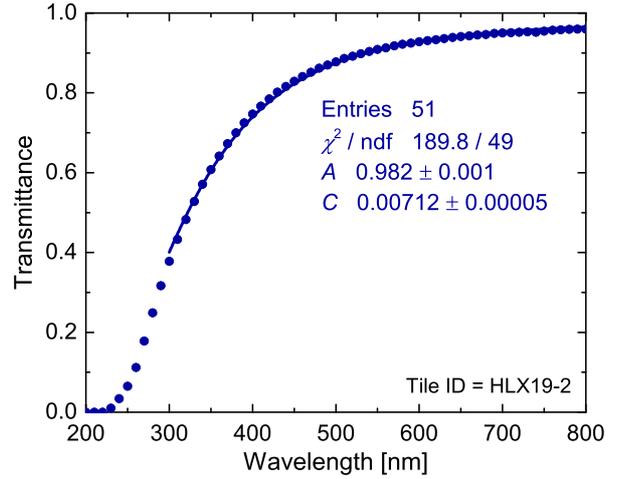}
\caption{UV--vis transmittance spectra recorded every 10 nm from the first mass-produced aerogel tile (circles). The transmittance ($T$) curve was fitted by using $T=A\exp(-C\cdot t/\lambda^4)$ (solid line), where $A$, $t$, and $\lambda $ are the amplitude, aerogel thickness, and wavelength of light, and $C$ is called the ``clarity coefficient.'' Data points ranging from 300 to 800 nm were fitted to obtain a better fitting result (i.e., small chi-squared) in the visible range. The parameters $A$ and $C$ were determined to be 0.982 and 0.00712 $\mu $m$^4$/cm.}
\label{fig:fig2}
\end{figure}

In addition to the absolute value, the uniformity of the refractive index and thickness distributions over a single monolithic tile must be predetermined for the precise measurement of $\beta $. The target precision of the velocity determination is $\Delta \beta /\beta < 0.1\%$. This resolution can be achieved by precisely mapping the index and thickness distributions of each tile with 0.1\% accuracy in advance. These mapped calibration data will be used in the offline analysis to determine $\beta $. It is helpful to reduce the number of mapping points on the raw aerogels that have a uniformity of approximately 1\% or better for both the index and thickness. A previous study on density uniformity of pin-dried aerogels showed rather negative results for the application of pin-dried aerogels as RICH radiators \cite{cite6}. However, subsequent studies significantly improved the density (i.e., refractive index) uniformity of pin-dried aerogels (data unpublished but see also Ref. \cite{cite7}). We expect our recent aerogels will meet the exact requirement.

We determined the realistic target transparency to be a transmission length of 30 mm or more at a wavelength of 400 nm after water-jet machining. Therefore, it is preferable for raw aerogels to have a transmission length of approximately 35 mm. The aerogel surface can be slightly damaged by aerogel powders generated by the water-jet cutting process and depending on how the aerogel is handled after cutting. Note that in principle water itself does not affect the aerogel optical properties due to a hydrophobic surface modification applied to the aerogel. The target transparency is comparable with the design of the downstream aerogel layer (a transmission length of 30 mm for $n$ = 1.055) of the ARICH detector in the Belle II spectrometer at Japan's High Energy Accelerator Research Organization (known as KEK) \cite{cite7}, which is a model of the HELIX RICH detector.

\section{Initial results from aerogel mass production}

We have previously reported a standard procedure for aerogel production without pin drying \cite{cite8}, and the concept of pin-drying wet-gel densification was also described in a previous paper \cite{cite3}. On a daily basis, four wet-gel tiles were synthesized at the same time, and after 1 h aging at 25$^\circ $C, they were treated by the pin-drying process. A group of four tiles that were synthesized on the same day are referred to as a ``lot.'' The pin drying of wet gels for approximately 70 days was then followed by surface modification using a hydrophobic reagent in addition to solvent exchange. Finally, drying by supercritical carbon dioxide was performed using supercritical fluid extraction equipment consisting of a 7.6 L autoclave. The autoclave could dry four lots (16 tiles) at once taking approximately 1 week, which was called a ``batch.''

\begin{figure}[t]
\centering 
\includegraphics[width=0.43\textwidth,keepaspectratio]{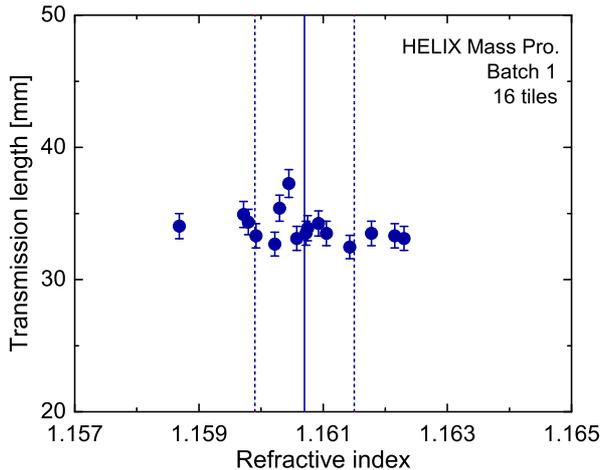}
\caption{Transmission length, calculated at $\lambda $ = 400 nm, as a function of the refractive index (in air). The data were taken from the 16 aerogel tiles produced in the first batch. The solid line shows the median. The dashed lines indicate the $\pm $0.5\% limits of the $n-1$ value.}
\label{fig:fig3}
\end{figure}

Laboratory-scale mass production of a total of 96 tiles was planned. The planned 96 tiles consisted of six batches, that is, 24 lots. By the end of July 2018, we had produced two batches, a total of 32 aerogel tiles. The wet-gel synthesis of the remaining 64 tiles was started at the beginning of July. By the beginning of October, another 16 aerogel tiles had been completed, and the remaining 48 wet-gel tiles were under hydrophobic treatment or pin drying.

The production of the first batch was completed on July 15, 2018. The first tile, taken as a typical example, measured 111 mm $\times $ 111 mm $\times $ 10.2 mm, with no cracking (Fig. \ref{fig:fig1}). The optical parameters were $n$ = 1.1597 (at 405 nm wavelength) and $\Lambda_{\rm T}$ = 35 mm, where $\Lambda_{\rm T}$ is the transmission length at 400 nm. This was one of the world's heaviest aerogel tiles having a density of 0.53 g/cm$^3$. Deviation of the index (+0.01) from the original design value ($n$ = 1.15) is acceptable, and it makes the cross-calibration with the ToF counters easier. An ultraviolet--visible (UV--vis) transmittance spectrum of the first tile along the 10.2 mm thickness direction is shown in Fig. \ref{fig:fig2}, which was used for calculating the transmission length. The methodologies for measuring the refractive index and transmission length can be found in literature \cite{cite8}.

Visual inspection confirmed a crack-free yield for the early batches of mass produced tiles. By the beginning of October 2018, we obtained 40 out of 48 tiles with no cracking (from three batches); thus, the crack-free yield was 83.3\%. This was close to our target level of crack-free tiles (90\%). The tile cracking occurred solely during the re-wetting process after pin drying. In the pin-drying process, the surface of the wet gels is partially dried. During the surface modification process by the hydrophobic reagent, the wet gels must be immersed in ethanol. This re-wetting process was performed step by step; however, tile cracking occasionally occurred because of this re-wetting process.

\begin{figure}[t]
\centering 
\includegraphics[width=0.48\textwidth,keepaspectratio]{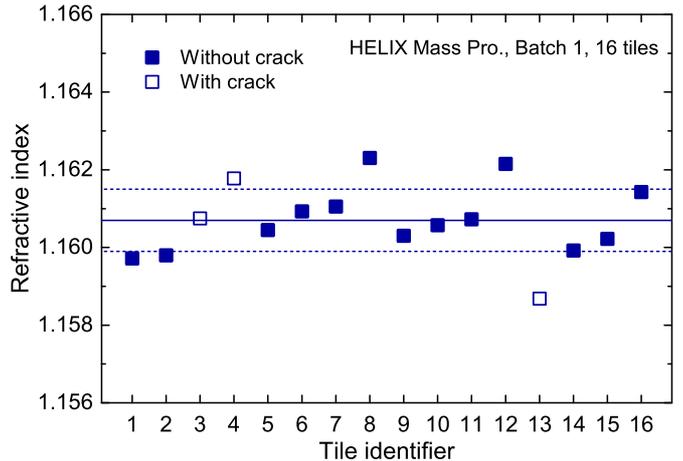}
\caption{Tagged refractive index, an average of the index values measured at the four corners of the 16 individual aerogel tiles produced in the first batch. The filled and open squares show tiles without and with cracking, respectively. The horizontal solid and dashed lines indicate the median and the $\pm $0.5\% limits of the $n-1$ value, respectively. A total of nine tiles were accepted as satisfying the index requirement with no cracking.}
\label{fig:fig4}
\end{figure}

Results from the basic optical characterizations performed just after production was promising for the first batch. Fig. \ref{fig:fig3} shows the transmission length as a function of the tagged refractive index (recorded with a laser in air), which is defined as the mean index measured at the four corners of individual tiles \cite{cite6}. The mean transmittance and transmission length were 73.7\% and 33.9 mm, respectively, which were consistent with our expectations. The median of the tagged index was 1.1607 (shown by the solid line in Fig. \ref{fig:fig3}). Imposing a limit of $\pm $0.5\% on the tagged index (specifically, $\pm $0.5\% of $n-1$), as shown by the dashed lines in Fig. \ref{fig:fig3}, 10 out of the 16 tiles fulfill this criterion. Among the 10 tiles, 1 tile was excluded because of tile cracking. Fig. \ref{fig:fig4} shows the tagged index for the first 16 tiles, which are divided into two groups, those with or without cracks. Therefore, 9 out of the 16 tiles satisfied the basic optical requirements.

\begin{figure}[t]
\centering 
\includegraphics[width=0.48\textwidth,keepaspectratio]{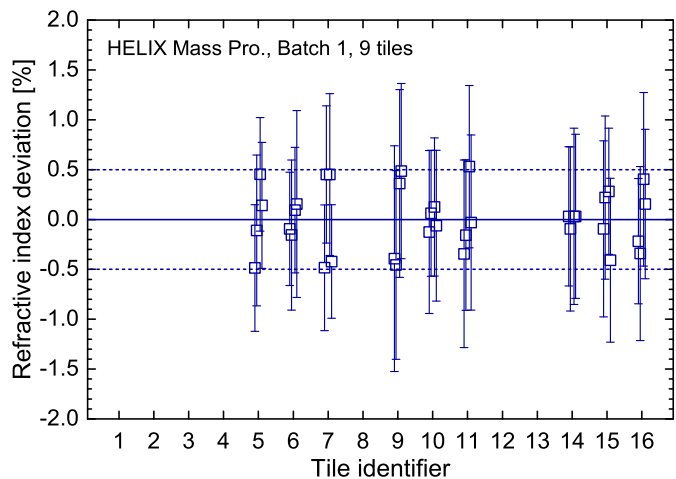}
\caption{Deviation from the mean refractive index was measured at the four individual corners of each tile. Data were taken from the nine tiles accepted in Fig. \ref{fig:fig4}. The dashed lines indicate the $\pm $0.5\% deviation limits of the $n-1$ value. For each tile, the data symbols (open squares) were slightly shifted along the $x$-axis to improve visibility.}
\label{fig:fig5}
\end{figure}

A total of eight tiles passed the first screening based on the basic optical measurement results, including a search for visual cracking, transmission length, refractive index at the four corners, and their mean value. For the nine tiles that were acceptable, Fig. \ref{fig:fig5} shows the deviation from the mean index of each tile. For individual tiles, the index was measured at the four corners of the tile, and the mean index was calculated by averaging the four sets of data concerning each tile. The mean index corresponds to the horizontal line at 0\% in Fig. \ref{fig:fig5}. Every index data measured at the four corners are shown as a function of the tile identification number. The dashed lines indicate the $\pm $0.5\% deviation limits of the $n-1$ value. Tile number 11 was rejected because of a slight deviation from +0.5\% for a single data point.

The selected eight tiles will be subjected to further investigations to check their uniformity. This includes checking the refractive index uniformity as well as the thickness uniformity over a single tile. Developing a method for measuring the uniformity is currently still in progress. Note that the screening results discussed in this section were based on the median/mean of the first 16 tiles and it should be reconsidered once the mass production of a full set of tiles has been completed.

\begin{figure}[t]
\centering 
\includegraphics[width=0.38\textwidth,keepaspectratio]{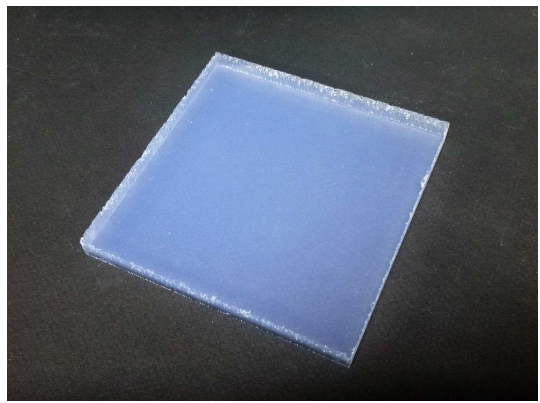}
\caption{Water-jet trimmed aerogel tile (tile sample was from an engineering production batch before mass production had begun).}
\label{fig:fig6}
\end{figure}

The trimming of the aerogel tiles by a water-jet cutting device from approximate dimensions of 116 mm $\times $ 116 mm to exact dimensions of 100 mm $\times $ 100 mm was tested successfully. This was the first water-jet machining of aerogels that have a refractive index above 1.12. Instead of the mass produced tiles, prototype tiles that had been fabricated before the mass production were used in the initial trimming test. The trimmed tile measured 99.5--99.75 mm, and no significant damage was visible, as shown in Fig. \ref{fig:fig6}. The tile had sufficient hydrophobic characteristics, and the cutting device was operated under precisely controlled conditions. Instead of the refractive index, the density of the tile was gravimetrically measured, and it was found to be consistent with the value measured before the machining. Therefore, we expect no significant change in the index based on the linear relationship between the $n-1$ value and density \cite{cite8}, which will be thoroughly investigated elsewhere using direct index measurement. Also, the transmittance degradation was below 1\% point at $\lambda $ = 400 nm and acceptable.

\section{Conclusions}

The HELIX RICH system employs highly transparent silica aerogel tiles with the highest refractive index ($n$ = 1.16) ever used as Cherenkov imaging radiators and will be employed in the upcoming first balloon flight in Antarctica. For this application, we mass produced hydrophobic aerogel tiles with dimensions of 11 cm $\times $ 11 cm $\times $ 1 cm. At present (as of October 2018), we have completed the fabrication of 48 aerogel tiles, obtaining 40 crack-free tiles. The basic optical properties of the first 16 tiles have been confirmed; the average transmittance was 73.7\% at $\lambda $ = 400 nm, which corresponds to a transmission length of 34 mm. A total of 8 out of the 16 tiles were selected as aerogels having good index uniformity in the first screening from the first batch. The first water-jet trimming test was successful. Mass production will finish by the end of November 2018.

\section*{Acknowledgments}

The wet silica gel production and optical measurements of the aerogel tiles were performed at the Venture Business Laboratory of Chiba University and the High Energy Accelerator Research Organization (KEK), Japan, respectively. The authors are thankful to them both for their support. Profs. I. Adachi of KEK and H. Kawai of Chiba University are thanked particularly for their help in aerogel fabrication and measurements. We are grateful to Tatsumi Kakou Co., Ltd., Japan, for their contributions to the water-jet machining. This study was supported by a Grant-in-Aid for Scientific Research (C) (No. 18K03666) from the Japan Society for the Promotion of Science (JSPS).



\end{document}